# Stimulated Brillouin scattering in chiral photonic crystal fiber

Xinglin Zeng,[1,*] Wenbin He,[1] Michael H. Frosz,[1] Andreas Geilen,[1] Paul Roth,[1] Gordon K. L. Wong,[1] Philip St.J. Russell,[1] and Birgit Stiller[1,2]

[1]*Max-Planck Institute for the Science of Light, Staudtstr. 2, 91058 Erlangen, Germany*
[2]*Department of Physics, Friedrich-Alexander-Universität, Staudtstr. 2, 91058 Erlangen, Germany*
*xinglin.zeng@mpl.mpg.de

**Abstract:** Stimulated Brillouin scattering (SBS) has many applications, for example, in sensing, microwave photonics and signal processing. Here we report the first experimental study of SBS in chiral photonic crystal fiber (PCF), which displays optical activity and robustly maintains circular polarization states against external perturbations. As a result, circularly polarized pump light is cleanly back-scattered into a Stokes signal with the orthogonal circular polarization state, as is required by angular momentum conservation. By comparison, untwisted PCF generates a Stokes signal with an unpredictable polarization state, owing to its high sensitivity to external perturbations. We use chiral PCF to realize a circularly polarized continuous-wave Brillouin laser. The results pave the way to a new generation of stable circularly polarized SBS systems with applications in quantum manipulation, optical tweezers, optical gyroscopes and fiber sensors.

## 1. Introduction

First observed in bulk materials by Chiao, Townes and Stoicheff in 1964 [1], stimulated Brillouin scattering (SBS) is a nonlinear optical process in which light is back-scattered by a hypersonic acoustic wave. The Doppler-shifted backward signal beats with the pump light, creating a moving interference pattern that in turn amplifies the acoustic wave, leading to strong amplification of the backward signal for high enough pump power. Since its first demonstration, SBS has been explored and exploited in many different systems, especially optical fibers [2] and integrated photonics [3]. Among the many applications of SBS are narrow linewidth lasers [4], fiber sensors [5], light storage systems [6] and microwave photonic filters [7]. To date, SBS has not yet been studied in chiral photonic crystal fibers, where the core microstructure rotates with position along the fiber axis.

Studies of SBS in non-chiral photonic crystal fibers (PCF) with high air-filling fractions and very small cores have revealed how tight confinement of acoustic vibrations gives rise to SBS frequency shifts not seen in standard step-index fiber [8] [9]. In recent years, chiral PCF, drawn from a spinning preform, has emerged as a unique platform for studying the behavior of light in chiral structures that are infinitely extended in the direction of the twist [10]; such structures are very difficult if not impossible to realize on an integrated photonic chip. Chiral PCF has been shown to robustly preserve circular polarization state over long distances, allowing investigation of nonlinear processes in the presence of chirality [11] [12]. Although the polarization properties of SBS in non-chiral fibers have been investigated by many research groups [13-16], it has so far not been possible to observe SBS between clean circularly polarized modes.

Here we report the first experimental study of SBS in chiral PCF, demonstrating both Brillouin amplification of circularly polarized light, and a continuous-wave (CW) circularly polarized Brillouin laser. The results are of potential interest in fiber optic current sensing [17], fiber-optic gyroscopes [18] and teleportation of quantum states [19].

## 2. Chiral photonic crystal fiber

Figure 1(a) shows a 3D sketch of chiral PCF. Figure 1(b) shows a scanning electron micrograph (SEM) of the microstructure of the chiral PCF used in the experiments. It was fabricated from fused silica using the standard stack-and-draw technique. The helical pitch was 1.6 cm, created by spinning the preform during the fiber draw, and the circular birefringence $B_C$ was measured to be 2.134 μRIU at 1550 nm (Fig. 1(c)), which agrees well with the results of finite element modelling (FEM) (for more details see Appendix A). The inset in Fig. 1(c) show the simulated distribution of the Poynting vector for the left-circularly polarized (LCP) fundamental mode Due to the non-degeneracy, the measured Stokes parameters are $(S_1, S_2, S_3) = (−0.27, 0.17, −0.95)$ and $(S_1, S_2, S_3) = (0.28, −0.18, 0.94)$ after the LCP and RCP modes propagate through 38 m chiral PCF, respectively.

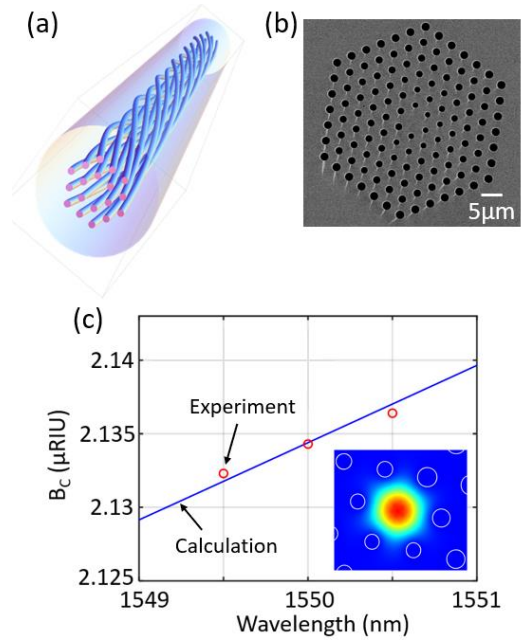

Fig. 1: (a) 3D sketch of a chiral PCF. (b) Scanning electron micrograph (SEM) of the PCF used in the experiments. The hollow channels have diameter $d = 1.7$ μm and spacing $\Lambda = 3.8$ μm ($d/\Lambda = 0.45$), and the core diameter is ~6 μm. The $LP_{01}$-like core modes have (measured) loss 0.06 dB/m (LCP) and 0.05 dB/m (RCP) and effective indices 1.435599 (LCP), and 1.435601 (RCP) at 1550 nm, yielding circular birefringence $B_C = 2.134$ μRIU. The Stokes $S_3$ parameters, measured after propagation along 38 m of fiber, were −0.95 (LCP) and +0.94 (RCP). (c) Measured and calculated circular birefringence $B_C$ versus wavelength. The inset is the calculated distribution of the Poynting vector.

## 3. Brillouin scattering in chiral PCF

Figure 2(a) shows the heterodyne detection setup for measuring Brillouin frequency in chiral PCF. Both pump and local oscillator (LO) are derived from a narrow linewidth (<1 kHz) 1550 nm continuous wave (CW) laser by using a 90:10 fiber coupler. The pump wave is then boosted by an Erbium-doped fiber amplifier (EDFA) and injected into the chiral PCF through an optical circulator. The circular polarization states are tuned by adjusting the fiber polarization controller (FPC) before the circulator. The thermal noise-initiated Stokes signal coming back from the PCF is delivered by the circulator and interferes with the LO by using the second 90:10 fiber coupler and the beating signal is detected by photodetector (PD) and electrical spectrum analyzer (ESA). Figure 2(b) shows the measured Brillouin spectrum in a 38 m length of chiral PCF for a pump power of 0.9 W, which is just below the threshold for SBS (the SBS threshold is usually defined as the point at which the Stokes power equals 1% of the pump power [20]). The peak at 10.82 GHz is an artefact caused by the Brillouin scattering in single-mode fiber after port 2 of the circulator, and the second peak at 11.013 GHz comes from the chiral PCF. Unlike PCFs with large $d/\Lambda$ values and µm-scale cores, which support several hybrid torsional-radial acoustic modes [8], the PCF used here has relatively low $d/\Lambda$ (0.436) and a large core (diameter 6 µm), so that the Brillouin frequency shift is close to that obtained in bulk glass: $2n_s c_L/\lambda_p$ = 11.12 GHz, where $n_s$ is the refractive index of silica at the pump laser wavelength $\lambda_p$, and $c_L$ = 5971 m/s is the longitudinal acoustic velocity.

To confirm these results, we used FEM to calculate the normalized optoacoustic coupling coefficient, defined by [21]:

$$\kappa = \left| \frac{\int\int E_{LC} E_{RC} p_{ijkl} \varepsilon_{kl} dxdy}{\int\int |E_{LC} E_{RC}| dxdy \int\int |\varepsilon|^2 dxdy} \right|^2, \quad (1)$$

where $E_{LC(RC)}$ are the scalar electric fields of circularly polarized pump and Stokes modes, $p_{ijkl}$ is the elasto-optic tensor and $\varepsilon_{kl}$ is the strain tensor associated with the acoustic mode. The numerical results in Fig. 2(c) show a strong peak at 11.02 GHz (close to the experimental peak at 11.013 GHz) and a weak subsidiary peak at 11.039 GHz. Although it was not possible to resolve these closely spaced peaks by heterodyning, they could be detected when the Stokes signal was seeded (see Section 4). The calculated power-normalized axial displacements of the two acoustic modes in Fig. 2(d) show clearly that the mode at 11.02 GHz is more concentrated in the PCF core and so is expected to be dominant.

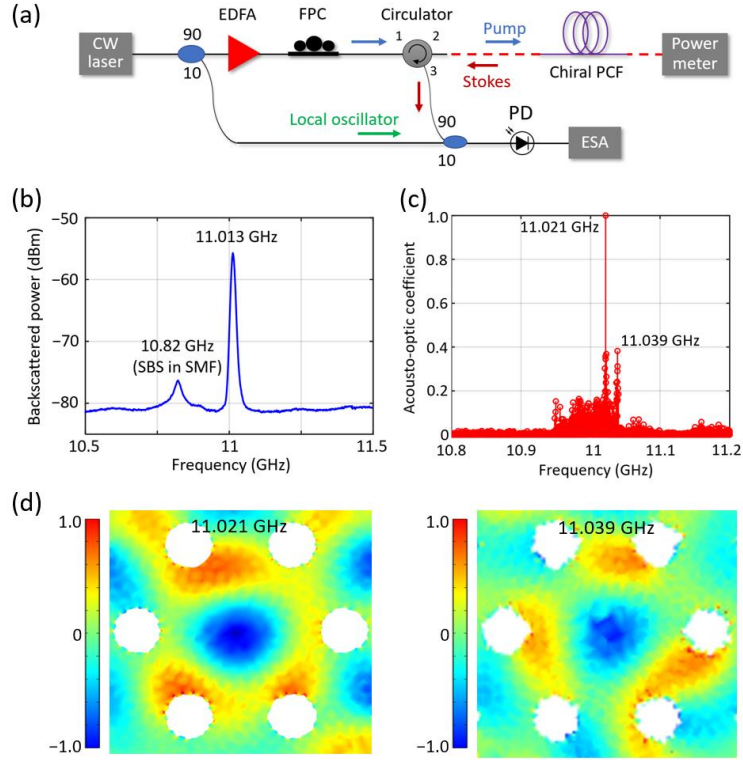

Fig. 2: (a) Experimental setup for measuring Brillouin frequencies in chiral PCF. (b) Brillouin spectrum generated by pumping with 0.9 W (just below the SBS threshold) of circularly polarized CW laser at 1550 nm. (c) Numerically calculated optoacoustic coupling coefficients (normalized) for different acoustic modes around 11 GHz. (d) Axial displacements (normalized to the square-root of the power) of acoustic modes at 11.021 GHz and 11.039 GHz in (c).

We next increased the pump power so as to reach the SBS regime, when the polarization state of the much stronger Stokes signals could be more easily and precisely measured (Fig. 3). The CW pump light was amplified in an erbium-doped fiber amplifier (EDFA) and its polarization state controlled using a combination of polarizing beam splitter (PBS) and quarter-wave (λ/4) plate. Back-scattered signals with the same polarization state as the pump are transmitted by the PBS and detected by a power meter placed at port 3 of the circulator, while orthogonally polarized light is reflected and detected by a second power meter. The transmitted pump power is monitored by a third power meter. Narrow-band (6 GHz) notch filters in the path of each Stokes signal are used to filter out Fresnel reflections and Rayleigh scattering. Fig. 3(b) shows the power dependence of the Stokes and transmitted pump signals for LCP, RCP and linearly polarized pump light. Above a threshold of ~0.95 W, the orthogonally polarized Stokes signal grows rapidly with a slope efficiency of ~50%, while the transmitted pump power saturates. The co-polarized Stokes signal shows no gain, as angular momentum conservation would otherwise be violated.

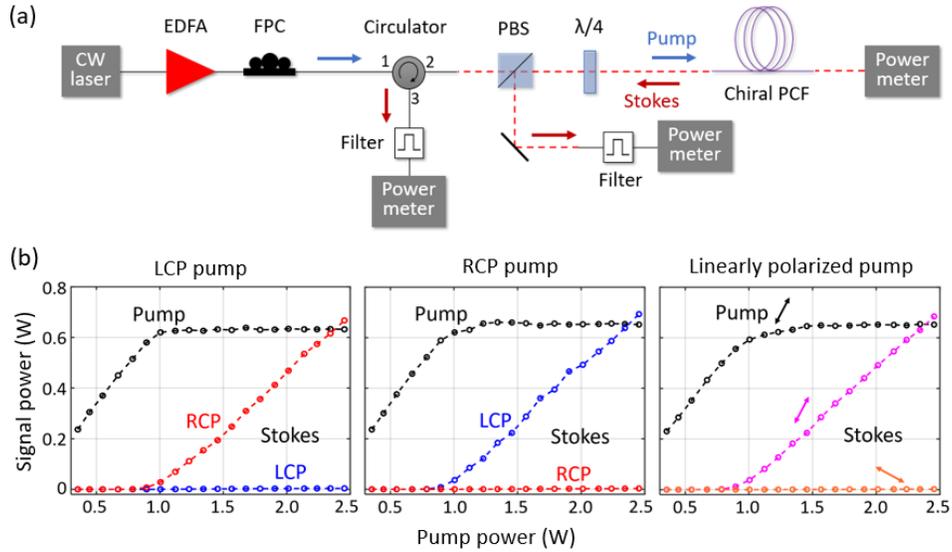

Fig. 3. (a) Experimental setup for measuring SBS threshold of different circularly polarized light in chiral PCF. EDFA: Erbium-doped fiber amplifier, FPC: fiber-based polarization controller, PBS: polarized beam splitter. (b) Stokes and transmitted pump power in a 38 m length of chiral PCF for LCP, RCP and linearly polarized pump light (left to right).

These results confirm robust maintenance of circular polarization states and conservation of spin during Brillouin scattering in chiral PCF. More interestingly, when a linearly polarized pump is injected into the chiral PCF, the polarization state of the backscattered Stokes wave is also linearly polarized and has same azimuthal angle as that of the pump, as shown in Fig. 3(b). We attribute this to well-controlled optical activity in the chiral PCF, which causes the linearly polarized pump and backward Stokes modes to be co-polarized at all points along the fiber. The SBS threshold power in all the three cases was ~0.95 W, i.e., it is independent of the polarization state.

To assess the robustness of polarization-maintenance, the signal powers and Stokes polarization states were measured for LCP and RCP pump light with the fiber spooled to diameters of 50 and 16 cm (Table 1). The pump power was kept constant at 2.45 W in all the measurements. The modulus of the Stokes parameter $S_3$:

$$|S_3| = \left| \frac{P_{RC} - P_{LC}}{P_{RC} + P_{LC}} \right|, \qquad (2)$$

was greater than 0.98 for both LCP and RCP pump light in the 50 cm spool, corresponding to an LCP to RCP power extinction ratio of better than 23 dB. For the 16 cm spool the $S_3$ was ~0.88, corresponding to a somewhat lower extinction ratio of 12 dB. Bending causes the guided eigenmodes to become elliptically polarized, which naturally affects the Brillouin signals. Due to the loss of pump and Stokes wave in the fiber, the input pump power is larger than the sum of total Stokes power and output pump power.

Table 1. Signal powers and Stokes $|S_3|$ in chiral PCF spooled to different diameters. Pump power is 2.45 W.

|  | LCP Stokes | | RCP Stokes | | Trans. pump | | $|S_3|$ | |
| --- | --- | --- | --- | --- | --- | --- | --- | --- |
| Spool diameter (cm) | 16 | 50 | 16 | 50 | 16 | 50 | 16 | 50 |
| LCP pump (mW) | 44 | 3 | 720 | 693 | 628 | 632 | 0.885 | 0.991 |
| RCP pump (mW) | 618 | 667 | 38 | 4 | 641 | 652 | 0.884 | 0.988 |

To illustrate the striking contrast between chiral and non-chiral PCFs, we carried out the same experiment using a 25 m length of untwisted PCF with a closely similar microstructure. The fiber loss was 0.012 dB/m and its slightly two-fold rotationally symmetric structure (Fig. 1(b)) resulted in a linear birefringence $B_L = 12.5$ μRIU at 1550 nm. Due to degeneracy between two circular polarization states, the measured Stokes parameters are $(S_1, S_2, S_3) = (−0.37, −0.74, −0.56)$ and $(S_1, S_2, S_3) = (0.11, −0.8, 0.58)$ after the LCP and RCP modes propagate through 25 m non-chiral PCF, respectively. Figure 4 shows the measured Stokes power as a function of input pump power. The shorter fiber length, and the poor polarization state maintenance, resulted in a twice higher SBS threshold power. In sharp contrast to chiral PCF, both the LCP and RCP Stokes signals grow in power together, which we attribute to the evolution of the pump polarization state, which in a straight fiber will go through a full cycle of LCP-linear-RCP-linear every $\lambda/B_L \sim 12$ cm.

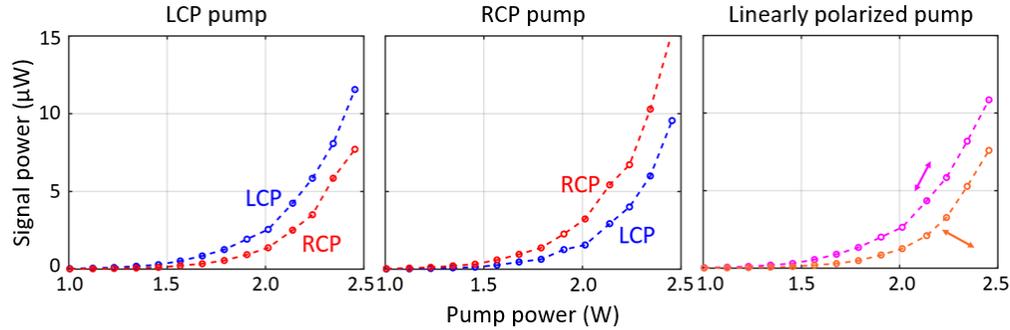

Fig. 4: Stokes signal powers measured in a 25 m length of untwisted PCF, pumped by LCP, RCP and linearly polarized light.

## 4. Brillouin amplification in chiral PCF

We now investigate using a 2 m length of chiral PCF as a Brillouin amplifier of circularly polarized light. The amplified Stokes signal takes the well-known form:

$$P_S(0) = P_S(L)\exp\left(g_B P_P(1-e^{-\alpha L})/\alpha - \alpha L\right) \qquad (3)$$

Where $g_B$ is the Brillouin gain coefficient, $\alpha$ the fiber loss, $L$ the fiber length and $P_P$ the pump power. Details of the experimental setup are available in the Appendix C. Figure 5(a) shows the Brillouin gain for varying pump-seed frequency difference at a pump power of 1 W and a Stokes seed power of 10 mW. As expected, the Brillouin gain is significant only when pump and seed are orthogonally polarized, reaching peak values of 0.82 W$^{-1}$m$^{-1}$ for LCP/RCP pump/seed and 0.9 W$^{-1}$m$^{-1}$ for the reverse. The gain spectra deviate slightly from perfect Lorentzians, but can be fitted to two Lorentzians, which we attribute to the two acoustic modes shown in Fig. 2(c). The double Lorentzian fits very well to the experimental results, with central frequencies 11.01 GHz and 11.044 GHz, in excellent agreement with the numerical calculations (11.02 GHz and 11.039 GHz).

The peak gain coefficients of the two Lorentzians are 0.745 $W^{-1}m^{-1}$ and 0.15 $W^{-1}m^{-1}$ for LCP pumping, and 0.794 $W^{-1}m^{-1}$ and 0.184 $W^{-1}m^{-1}$ for RCP pumping, again close to the theoretical values of 0.851 $W^{-1}m^{-1}$ and 0.256 $W^{-1}m^{-1}$ respectively (for details see Appendix B). The linewidth of fitted Lorentzian gain spectrum is 31 MHz for the stronger acoustic mode and 65 MHz for the weaker one, yielding an effective gain linewidth of 41 MHz for both pumping configurations. The gain at four different LCP pump powers is plotted versus frequency in the left-hand panel in Fig. 5(b), and versus pump power in the right-hand panel, showing a slope efficiency of 5.67 dB/W.

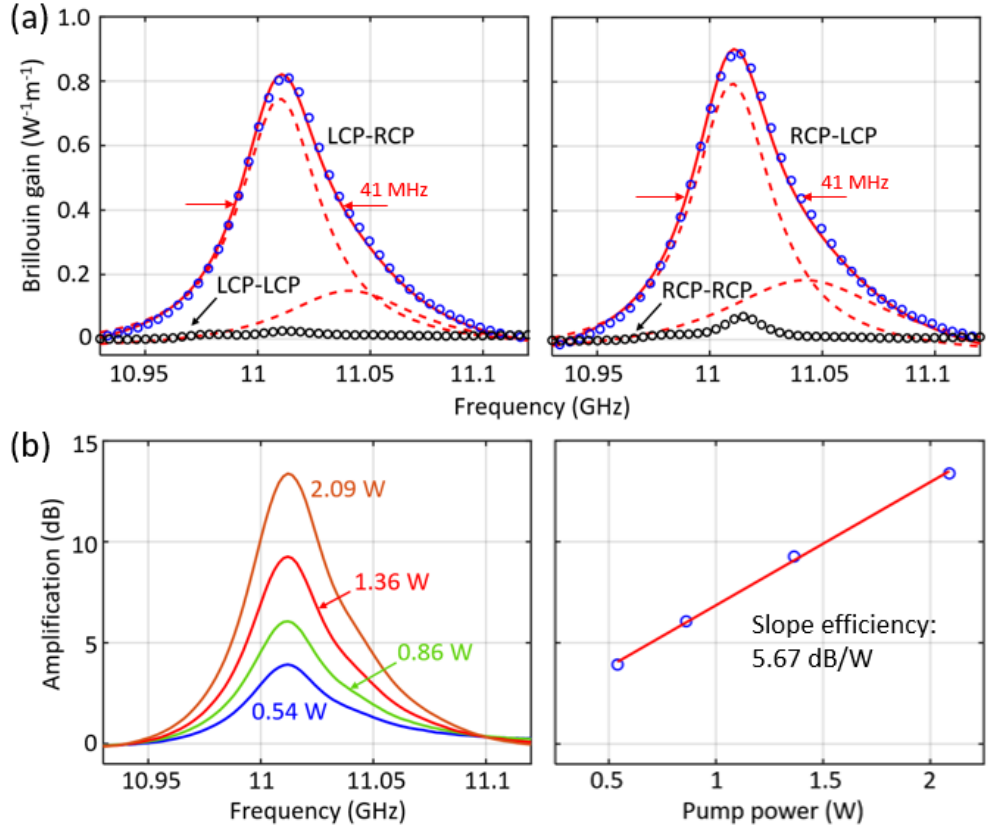

Fig. 5: (a) Measured Brillouin gain spectra (FWHM 41 MHz) when pumping with LCP (left) and RCP (right) light. The circles are measured data, and the full red line is a fit based on two Lorentzians (FWHM 31 and 65 MHz), indicated by the dashed red lines. (b) Left: Stokes gain spectra for LCP pump powers 0.54 W, 0.86 W, 1.36 W and 2.09 W. Right: Peak Stokes gain as a function of LCP pump power. The circles are the experimental datapoints and the line is a linear fit.

## 5. Circularly polarized Brillouin laser

Next, we constructed a circularly polarized Brillouin laser by placing a 2 m length of chiral PCF in a ring cavity (Fig. 6(a)). CW pump light was launched into the laser cavity through a polarizing beam-splitter and a λ/4 plate, so as to generate a circularly polarized signal. Since the Stokes signal is orthogonal to the pump it is reflected by the polarizing beam-splitter and thus circulates inside the cavity. A λ/4 plate and a polarizer are used to block the transmitted pump light, while letting the backward Stokes signal propagate freely. A beam-splitter is used to couple 10% of the Stokes signal out of the cavity, so that its polarization state can be measured. The laser spectrum, threshold and linewidth were measured by an optical spectrum analyzer (OSA), a power meter and by self-heterodyning. The total cavity length was 4 m (2 m chiral PCF and 2 m free-space), yielding a free spectral range of 30.8 MHz, which given the 41 MHz FWHM Brillouin gain linewidth means that only one cavity mode participates in lasing.

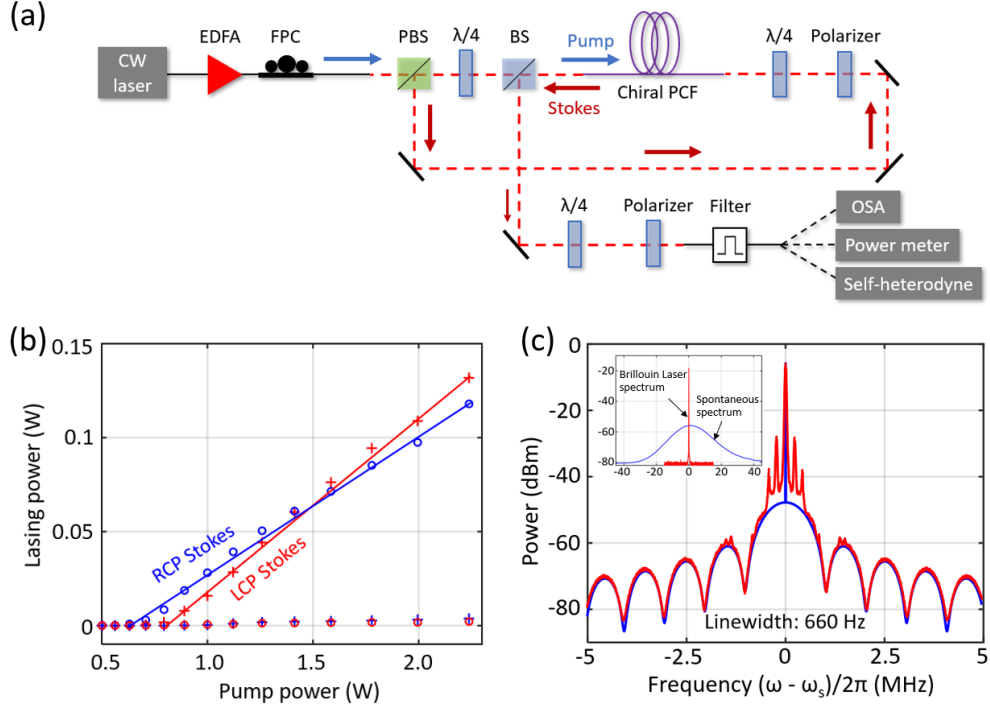

Fig. 6: (a) Experimental setup of circularly polarized Brillouin laser. (b) Laser power as a function of pump power for LCP (blue circles) and RCP (red crosses) pump light. (c) Theoretically fitted (blue) and measured (red) laser spectrum from sub-coherence delayed self-heterodyne system. Four sharp side-peaks around the center peak is from the mechanical modulation of pump laser and unrelated to the experimental result. The inset is the spectra of spontaneously scattered Stokes light from the identical chiral PCF and the linewidth-narrowed intracavity laser spectrum measured by delayed self-heterodyne system (the axis labels are same as those of main figure).

The spectrum measured by the OSA just before the filter confirmed the presence of a lasing signal with frequency 11.013 GHz below the pump frequency (for details see Appendix D). The laser power is plotted against LCP and RCP pump power in Fig. 6(b). Lasing commences when the Brillouin gain exceeds the 4.12 dB round-trip loss of the laser cavity (beam splitter 1 dB; fiber coupling, 1.5 dB; fiber loss 0.12 dB; polarization components 1.5 dB). The threshold powers are 708 mW for LCP pumping (slope efficiency 5.8%) and 794 mW for RCP pumping (slope efficiency 7.6%). The Brillouin laser spectrum measured by delayed self-heterodyning [22] (for details see Appendix E) is shown in Fig. 6(c). The laser spectrum is also compared in the inset of Fig. 6(c) with the spontaneous Stokes spectrum generated in identical chiral PCF in the absence of optical cavity feedback, showing a line-narrowing factor of ~$5\times10^4$, from ~31 MHz to 660 Hz.

We note that lasers with sub-kHz linewidth have a coherence length of great than 100 km, and the decoherent self-heterodyne requires the SMF in reference path to be longer than that. Such a long fiber transmission will induce Gaussian noise on the laser spectrum. So as to precisely measure the Brillouin laser linewidth we applied a "sub-coherence" self-heterodyne technique using a 2 km length of single-mode fiber as a delay line, so that the path difference was less than the laser coherence length and the laser spectrum has less Gaussian noise. The exact laser linewidth of 660 Hz is then obtained by fitting the measured spectrum to the sub-coherence lineshape function, as shown in the Fig. 6(c).

## 6. Conclusions

Chiral PCF, drawn from a spinning preform, robustly maintains the spin of guided light and so may be used for stimulated Brillouin scattering with circularly polarized light. Conservation of angular momentum means that significant gain is only possible when the Stokes and pump signals are orthogonally circularly polarized. Brillouin amplifiers and lasers for circularly polarized light can be successfully realized using chiral PCF as the gain medium. When, in contrast, non-chiral PCF is used, the polarization states of the Stokes and transmitted pump signals are unpredictable. Chiral SBS has many potential applications, for example in fiber sensing [17] [23], optical tweezers [24], Brillouin laser gyroscopes [25], and quantum manipulation [19]. Chiral N-fold rotationally symmetric multicore fibers support circularly polarized modes that carry orbital angular momentum, and can also be used in SBS experiments [26]. Finally, we note that long chiral waveguiding structures are extremely difficult if not impossible to realize on integrated photonic chips, making chiral PCF a unique vehicle for exploring Brillouin scattering in twisted structures.

## Appendix A: Optical and acoustic modes simulation for chiral PCF

The geometric model of chiral photonic crystal fiber (PCF) used for finite element method (FEM) simulation is based on a scanning electron microscopy (SEM) image of the fiber. Fig. 7(a) shows the SEM image of the chiral PCF. In order to make the calculation more precisely, a refined mesh system is configured in the core area. The fiber structure is slightly asymmetric and the core diameter is 6.16 μm for the major axis and 5.86 μm for the minor axis. The air-hole diameter d is ~1.7 μm and inter-hole spacing Λ is ~3.8 μm. The d/Λ is 0.45, which close to the endlessly single-mode condition. The fiber twist pitch is 1.6 cm.

A non-orthogonal coordinate transformation between helicoidal frame and Cartesian coordinate has to be applied for optical simulation with chiral PCF. The matrix transformation is detailed in Ref. [10] and the new permittivity and permeability tensors in chiral PCF can be obtained by simply inducing this transfer matrix on the original material permittivity and permeability in the straight PCF. After implementing this step in the Wave Optics module of COMSOL, the Maxwell equations can be solved numerically with FEM. Figures 7(b) shows the simulated mode profiles and refractive indices of left circular polarization and right circular polarization states. The effective mode areas for both modes are 24.6 μm$^2$.

Figure 7(c) show calculated axial displacements of acoustic modes that have the highest overlapping coefficients with the optical modes, by using the Solid Mechanics module of COMSOL. The acoustic frequencies are 11.021 GHz and 11.039 GHz. Due to the low d/Λ of this PCF, both acoustic modes are not completed confined in the fiber core region, but spread into the fiber structure. This is basically the same as the results of forward Brillouin scattering in single-mode fiber (SMF), where acoustic energy fills the whole cladding and a higher-order acoustic mode has the largest acousto-optic overlap with the fundamental optical mode. Here, two acoustic modes have symmetric vibration in the fiber core region and thus have no difference in both twisted and untwisted cases.

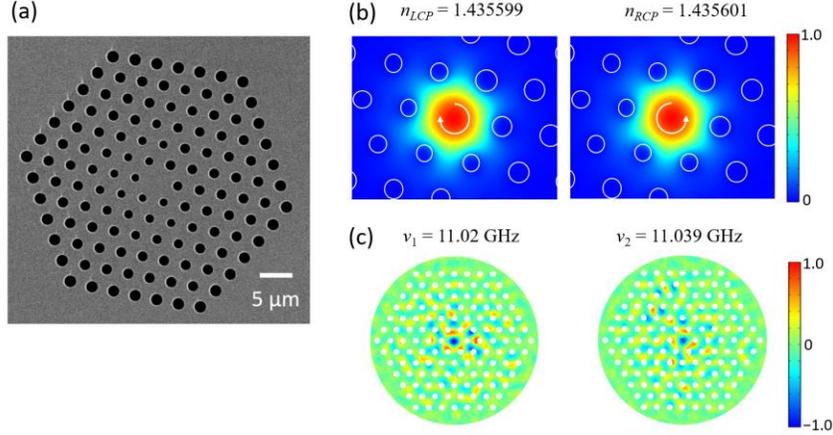

Fig. 7: SEM of cross-section of the chiral PCF. (b) Calculated normalized electric fields and effective indices for LCP and RCP modes. (c) Calculated axial displacements and frequencies for two acoustic modes that contribute to the backward Brillouin scattering in chiral PCF

## Appendix B: Theoretical estimation of Brillouin gain in chiral PCF

We then roughly estimate the gain coefficient by using equation [2]:

$$g_0 = \frac{\omega \gamma_e^2}{n_{\text{eff}} v_a c^3 \rho \Gamma_B A_{\text{eff}}^{\text{ao}}}, \quad (A1)$$

where $\omega$ is the light angular frequency, $\gamma_e$ is the electrostrictive constant, $n_{\text{eff}}$ is the effective index of the optical mode in the chiral PCF, c is the optical velocity in vacuum, $\rho$ is the mean density of silica, $v_a$ is the longitudinal acoustic velocity and $\Gamma_B/2\pi$ is the gain spectrum linewidth. For silica, $\gamma_e$ =1.17, $\rho_0$ =2202 kg/m$^3$, $n_{\text{eff}}$ =1.4356 and $v_a$ = 5970 m/s. In the experiment, $\omega$=2π×193.414 THz, $\Gamma_B$=2π×31 MHz for acoustic mode at 11.01 GHz and $\Gamma_B$=2π×65 MHz for acoustic mode at 11.044 GHz. $A_{\text{eff}}^{\text{ao}}$ is the acousto-optic overlap effective area and defined as:

$$A_{\text{eff}}^{\text{ao}} = \left[\frac{\langle f^2(x,y)\rangle}{\langle \xi_m(x,y) f^2(x,y)\rangle}\right]^2 \langle \xi_m^2(x,y)\rangle, \quad (A2)$$

Where the $f^2(x,y)$ is the transverse optical intensity and $\xi_m(x,y)$ is the acoustic wave (longitudinal acoustic mode). The brackets $\langle..\rangle$ represent the overlap integral over the fiber cross-section. The $A_{\text{eff}}^{\text{ao}}$ are calculated as 24 μm$^2$ and 38 μm$^2$ for the two acoustic modes in Fig. 7(c). By inserting all the parameters into Eq. (A1), we can calculate the theoretical Brillouin gain coefficient to be 0.851 W$^{-1}$m$^{-1}$ and 0.256 W$^{-1}$m$^{-1}$ for two acoustic modes, which are very close to the measured Brillouin gain shown in the main manuscript (0.745 W$^{-1}$m$^{-1}$ and 0.15 W$^{-1}$m$^{-1}$ for pumping LCP, 0.794 W$^{-1}$m$^{-1}$ and 0.184 W$^{-1}$m$^{-1}$ for pumping RCP).

## Appendix C: Brillouin amplification and gain coefficient measurement setup

The Brillouin gain coefficient and spectra are obtained by using a typical pump-seed measurement setup with polarization control, as shown in Fig. 8. Both pump and seed are derived from a narrow linewidth 1550 nm CW laser, the seed light is frequency tuned using a single side-band modulator (SSBM). The pump signal is boosted by an EDFA and the polarization states of both pump and seed are controlled by fiber polarization controllers (FPCs). The polarization states are measured by a polarimeter. After propagating backwards through the chiral PCF, the seed signal is amplified and then measured by a power meter after being delivered by port 3 of circulator and a filter.

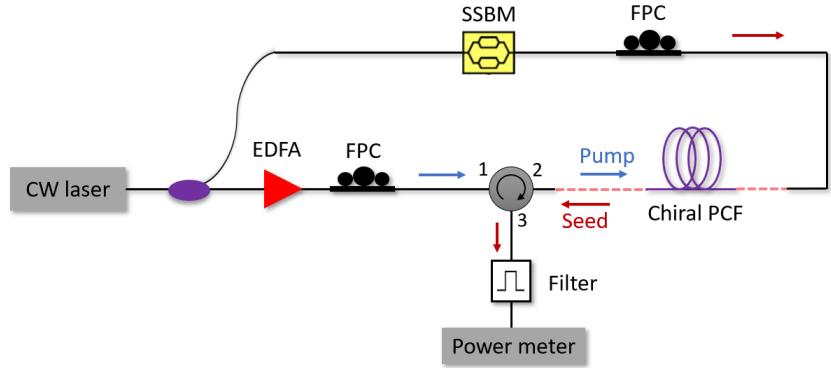

Fig. 8: Experimental setup for measuring Brillouin gain spectra of chiral PCF. (Dashed line represents free space)

## Appendix D: Brillouin laser spectrum measured by OSA

The spectrum (Fig. 9) measured by the optical spectrum analyzer (OSA) just before the narrow-band filter confirmed the presence of a lasing signal with frequency 11.013 GHz below the pump frequency.

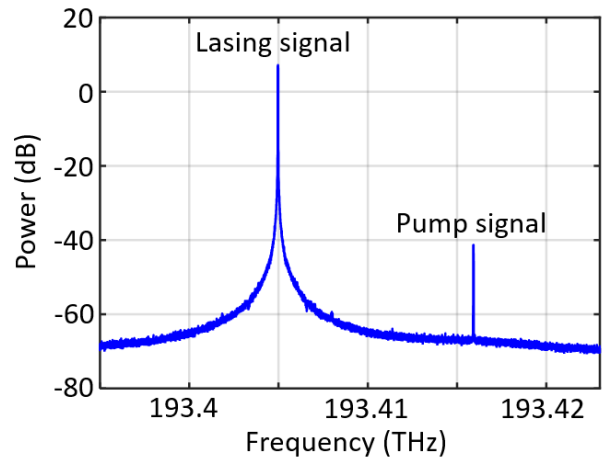

Fig. 9: Power spectrum of lasing light before the narrow-band filter measured by the OSA. The lasing signal and back-reflected pump signal are clearly observed.

## Appendix E: Delayed self-heterodyne setup for linewidth measurement

In order to precisely measure the linewidth of circularly polarized Brillouin laser, we use the delayed self-heterodyne interference technique for linewidth measurement. This technique is basically a Mach-Zehnder interferometer. It should be noted that simple measurement methods such as grating based OSAs or Fabry-Perot interferometers typically have insufficient resolution to measure the expected linewidth in this experiment. OSAs typically have a maximum resolution of a few GHz and Fabry-Perot interferometers can typically only measure down to tens of MHz. On the other hand, the delayed self-heterodyne interference technique can measure linewidths down to sub-kHz and hence was extensively employed for typical linewidth measurements.

Figure 10(a) shows the experimental setup. The output light from the Brillouin laser is coupled into a SMF and split into two paths by using a 3dB fiber coupler. One path is frequency shifted using an acousto-optic modulator (AOM) with 200 MHz modulation frequency, and the other path is transmitted through a long SMF, so that it becomes uncorrelated or sub-correlated (sub-coherent) with the frequency-shifted path. The two beams are then combined and interfere on a photodetector and the beat frequency (centered at 200 MHz) is observed on an ESA. A polarization controller is used in one of the arms to maximize the intensity of the beating signal.

The analysis of the resulting spectrum falls in two regimes: one for lasers with a coherent length shorter than or comparable to the SMF delay, and the other for lasers with coherent length longer than the SMF delay (the sub-coherent domain). Ideally, the laser light with a coherent length shorter than the imbalance delay will produce a Lorentzian spectrum with a half-width at half-maximum equal to the laser linewidth. However, due to a substantial contribution from Gaussian noise (i.e., pump noise, vibrations and acoustic noise) after long SMF transmission, the resulted spectrum profile is a convolution of Gaussian and Lorentzian functions and has a complex Voigt profile, as shown in Fig. 10(b) (the measured Brillouin laser spectrum with 26 km imbalance delay). Therefore, in this work, we use only several kilometers SMF delay to measure the linewidth in the sub-coherent range, and in that case the self-heterodyne lineshape function [27] will be the form of Dirac delta function at the AOM frequency combined with the interferometer transfer function that has infinite ripples away from center frequency. Figure 10(c) shows measured power spectrum using sub-coherent delayed self-heterodyne technique when SMF delay length is 2 km and 3 km, respectively. The side peaks near the central delta function is due to the thermal modulational noise of the pump laser. The linewidth of approximately 0.6 kHz is obtained by fitting the measured data with self-heterodyne lineshape function.

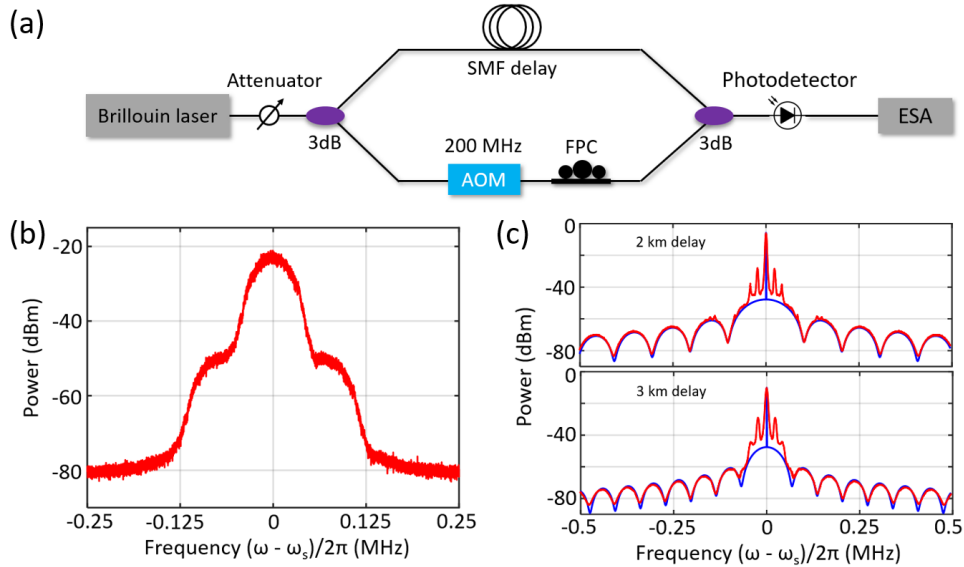

Fig. 10: (a) Experimental setup for delayed self-heterodyne measurement. (b) Experimentally measured self-heterodyne spectrum when the SMF delay is 26 km. (c) Experimentally measured self-heterodyne spectrum (red) and fitted curves (blue) with sub-coherent heterodyne formula [27] when the SMF delay is 2 km and 3 km.

**Funding.** Max-Planck-Gesellschaft (MPG).

**Acknowledgments.** The authors thank Yang Chen, Jennifer Hartwigs and Jiapeng Huang for help with several aspects of theory and experiments.

**Disclosures.** The authors declare no conflicts of interest.

**References**

1. R. Y. Chiao, C. H. Townes, and B. P. Stoicheff, "Stimulated Brillouin scattering and coherent generation of intense hypersonic waves," Phys. Rev. Lett. **12**, 592–595 (1964).
2. A. Kobyakov, M. Sauer, and D. Chowdhury, "Stimulated Brillouin scattering in optical fibers," Advances in Optics and Photonics **2**, 1–59 (2010).
3. B. J. Eggleton, C. G. Poulton, P. T. Rakich, Michael. J. Steel, and G. Bahl, "Brillouin integrated photonics," Nat. Photonics **13**, 664–677 (2019).
4. S. P. Smith, F. Zarinetchi, and S. Ezekiel, "Narrow-linewidth stimulated Brillouin fiber laser and applications," Opt. Lett. **16**, 393 (1991).
5. X. Bao and L. Chen, "Recent Progress in Brillouin Scattering Based Fiber Sensors," Sensors **11**, 4152–4187 (2011).
6. B. Stiller, M. Merklein, C. Wolff, K. Vu, P. Ma, S. J. Madden, and B. J. Eggleton, "Coherently refreshing hypersonic phonons for light storage," Optica **7**, 492 (2020).
7. A. Zarifi, M. Merklein, Y. Liu, A. Choudhary, B. J. Eggleton, and B. Corcoran, "Wide-range optical carrier recovery via broadened Brillouin filters," Opt. Lett. **46**, 166 (2021).
8. P. Dainese, P. St.J. Russell, N. Y. Joly, J. C. Knight, G. S. Wiederhecker, H. L. Fragnito, V. Laude, and A. Khelif, "Stimulated Brillouin scattering from multi-GHz-guided acoustic phonons in nanostructured photonic crystal fibres," Nat. Phys. **2**, 388–392 (2006).
9. J. C. Beugnot, T. Sylvestre, D. Alasia, H. Maillotte, V. Laude, A. Monteville, L. Provino, N. Traynor, S. F. Mafang, and L. Thevenaz, "Complete experimental characterization of stimulated Brillouin scattering in photonic crystal fiber," Opt. Exp. **15**, 15517–15522 (2007).
10. P. St.J. Russell, R. Beravat, and G. K. L. Wong, "Helically twisted photonic crystal fibres," Phil. Trans. R. Soc. A **375**, 20150440 (2017).
11. R. P. Sopalla, G. K. L. Wong, N. Y. Joly, M. H. Frosz, X. Jiang, G. Ahmed, and P. St.J. Russell, "Generation of broadband circularly polarized supercontinuum light in twisted photonic crystal fibers," Opt. Lett. **44**, 3964–3967 (2019).
12. S. Davtyan, D. Novoa, Y. Chen, M. H. Frosz, and P. St.J. Russell, "Polarization-tailored Raman frequency conversion in chiral gas-filled hollow-core photonic crystal fibers," Phys. Rev. Lett. **122**, 608 (2019).
13. M. O. van Deventer and A. J. Boot, "Polarization properties of stimulated Brillouin scattering in single-mode fibers," J. Lightwave Technol. **12**, 585–590 (1994).
14. D. Williams, X. Bao, and L. Chen, "Effects of polarization on stimulated Brillouin scattering in a birefringent optical fiber," Photon. Res. **2**, 126 (2014).
15. Kwang Yong Song, Weiwen Zou, Zuyuan He, and Kazuo Hotate, "All-optical dynamic grating generation based on Brillouin scattering in polarization-maintaining fiber," Opt. Lett. **33**, 926 (2008).
16. G. Prabhakar, X. Liu, J. Demas, P. Gregg, and S. Ramachandran, "Phase Conjugation in OAM fiber modes via Stimulated Brillouin Scattering," in *Conference on Lasers and Electro-Optics* (OSA, 2018), paper FTh1M.4.
17. R. Beravat, G. K. L. Wong, X. M. Xi, M. H. Frosz, and P. St.J. Russell, "Current sensing using circularly birefringent twisted solid-core photonic crystal fiber," Opt. Lett. **41**, 1672–1675 (2016).
18. V. M. N. Passaro, A. Cuccovillo, L. Vaiani, M. De Carlo, and C. E. Campanella, "Gyroscope Technology and Applications: A Review in the Industrial Perspective," Sensors **17**, 2284 (2017).
19. K. Tsurumoto, R. Kuroiwa, H. Kano, Y. Sekiguchi, and H. Kosaka, "Quantum teleportation-based state transfer of photon polarization into a carbon spin in diamond," Commun Phys **2**, 74 (2019).
20. R. W. Boyd, *Nonlinear Optics* (Academic Press, 2008).
21. V. Laude, A. Khelif, S. Benchbane, M. Wilm, T. Sylvestre, B. Kibler, A. Mussot, J. M. Dudley, and H. Maillotte, "Phononic band-gap guidance of acoustic modes in photonic crystal fibers," Phys. Rev. B **71**, 045107 (2005).
22. T. Okoshi, K. Kikuchi, and A. Nakayama, "Novel method for high resolution measurement of laser output spectrum," Electron. Lett. **16**, 630 (1980).
23. A. Küng, P.-A. Nicati, and P. A. Robert, "Brillouin Fiber Optic Current Sensor," in *Optical Fiber Sensors* (OSA, 1996), p. We21.
24. M. E. J. Friese, T. A. Nieminen, N. R. Heckenberg, and H. Rubinsztein-Dunlop, "Optical alignment and spinning of laser-trapped microscopic particles," Nature **394**, 348–350 (1998).
25. S. Huang, L. Thevenaz, K. Toyama, B. Y. Kim, and H. J. Shaw, "Optical Kerr-effect in fiber-optic Brillouin ring laser gyroscopes," IEEE Photon. Technol. Lett. **5**, 365–367 (1993).
26. X. Zeng, W. He, J. Huang, P. Roth, M. H. Frosz, G. K. L. Wong, B. Stiller, and P. St.J. Russell, "Stimulated Brillouin scattering of helical Bloch modes in 3-fold rotationally symmetric chiral 4-core photonic crystal fibre," in CLEO/Europe, Munich, Germany, paper CD-6.4, (2021).

27. L. Richter, H. Mandelberg, M. Kruger and P. McGrath, "Linewidth determination from self-heterodyne measurements with subcoherence delay times," IEEE J. Quantum Electron. **22**, 2070 (1986).